\documentstyle[12pt]{article}

\begin{document}
\def\la{\mathrel{\mathpalette\fun <}}
\def\ga{\mathrel{\mathpalette\fun >}}
\def\fun#1#2{\lower3.6pt\vbox{\baselineskip0pt\lineskip.9pt
        \ialign{$\mathsurround=0pt#1\hfill##\hfil$\crcr#2\crcr\sim\crcr}}}
\newcommand {\eegg}{e^+e^-\gamma\gamma~+\not \! \!{E}_T}
\newcommand {\mumugg}{\mu^+\mu^-\gamma\gamma~+\not \! \!{E}_T}
\renewcommand{\thefootnote}{\fnsymbol{footnote}}
\bibliographystyle{unsrt}

\begin{flushright}
UMD-PP-98-112\\
SMU-98-03
\end{flushright}
\begin{center}

{\Large \bf Nucleosynthesis constraints on massive, stable, 
strongly interacting particles}

\vskip8mm
{\bf  Rabindra N. Mohapatra$^{1}$ and Vigdor L. Teplitz$^{2}$}

\vskip6mm

{\it$^{(1)}${ Department of Physics, University of Maryland, 
College Park, Md-20742, USA.}}

{\it $^{(2)}${Department of Physics, Southern Methodist University,
Dallas, Tx-75275, USA}}

\end{center}
\begin{abstract}

We find constraints on heavy, stable, strongly interacting massive 
particles (X) from searches for anomalous nuclei containing them, formed 
during primordial nucleosynthesis. Using existing data, we obtain a limit 
on the abundance ratio $C_X\equiv n_X/n_B$ in the range of $3\times 
10^{-8}$ to $3\times 10^{-13}$ for masses up to 10 TeV if the $X-N$ 
interaction is sufficiently strong to bind in low Z nuclei.  
We also find a rough lower limit on the $X-N$ interaction 
that implies binding in nuclei with $A\geq 200$ over much of the $M_X$ 
range of interest, and address the relative abundance of such anomalous 
nuclei on Earth.

\end{abstract}
 
\newpage
\renewcommand{\thefootnote}{\arabic{footnote})}
\setcounter{footnote}{0}
\addtocounter{page}{-1}
\baselineskip=24pt

 \vskip0.5cm
Recently a number of authors have entertained the possibility 
that strongly interacting massive particles (SIMPS), with
masses a few GeV and above, may play a role in particle physics and 
cosmology\cite{starkman,chung,bere,nardi,moh,nandi,riotto}. For example, 
these
particles have been proposed as the source of ultra high energy cosmic rays
(UHECR) by Chung et al. and others\cite{chung,bere}. There are interesting 
models of
supersymmetry breaking where such particles appear as part of the 
messenger sector or as the gluino LSP\cite{nandi}. 

 Due to their 
strong interactions and high mass, these particles have distinct
consequences in cosmology and astrophysics. In order to study their 
implications, it is essential to know their annihilation cross section , 
$\sigma_{X\bar{X}}$ which in turn determines their relic density. It 
has recently been noted \cite{moh} that, if we assume $\sigma_{X\bar{X}}
\sim M^{-2}_X$, they can appear in 
sufficient abundance to saturate either the cosmic density or 
the galactic halo. Simple estimates of their relic density give
\begin{eqnarray}
\eta_X\equiv \frac{n_X}{n_{\gamma}} \simeq 5\times 10^{3}\frac{M_X}{M_{Pl}}
\end{eqnarray}
For a 1 TeV $X$ particle, this gives the relative abundance $C_X\equiv 
\frac{n_X}{n_B}\simeq 5\times 10^{-3}$
or $n_X\simeq 5\times 10^{-10}/cm^{3}$. For it to saturate the cosmic 
density, its mass must satisfy $M_X\geq 10^{3.5}$ GeV.  

 If in the process of cosmological
evolution, the $X$ particles formed the gravitational potential well and 
thus became the halo dark matter, their halo number density would be 
considerably  enhanced and become $n_X\simeq 3\times 10^{-4}/cm^3$ or so 
and known estimates of their density could be used to constrain their 
masses as functions of cross sections\cite{starkman,moh}.

In \cite{moh}, using the above 
form for the $X$ annihilation cross section as well as a simple QCD 
inspired ansatz for the $X-A$ cross it was shown 
that if the $X$'s constituted the halo dark matter and 
their masses were above 100 GeV, they would be in conflict with the results
from underground detectors looking for dark matter. The basic reason for 
this conclusion is that heavy nonrelativistic particles lose only a small 
fraction of their energy in elastic scattering and will therefore have to 
undergo a large number of collisions before they can slow down and get 
captured. Thus they should easily penetrate to reach the 
underground detectors but their signals have not been seen. There are 
also constraints\cite{moh} from the cosmological grand unified photon 
spectrum\cite{kolb} arising from particle-anti-particle annihilation if 
one assumes either both appear in equal abundance or the $X$ 
particles 
are their own anti particles. Again this would rule out $m_X\geq 100$ GeV
if one assumes the abundance to be halo density. Similar conclusions were
also derived using the limits on anomalous heavy isotope abundances. 

There have been several 
experimental searches \cite{rich,mcguire} to see whether these particles 
exist as dark matter and stringent limits have been
placed on their masses and their interaction cross sections with matter.
Other information on the possible existence of SIMPs used in 
both References\cite{starkman} and \cite{moh} comes from terrestrial 
searches for anomalous
heavy nuclei such as those in Ref. \cite{hemmick}. Specifically, the
experiment of Hemmick et al \cite{hemmick} set stringent upper limits on
the abundance of nuclei containing such heavy stable particles. These
limits were used in Ref.'s\cite{starkman,moh} to exclude SIMPs above 
a mass of 1-10 TeV. In order to study this question, one needs to know
the strength of $X-N$ binding potential $V_{XN}$, which is apriori 
unknown. One may either leave it as a free parameter as we do in the 
first part or use a ``factorization''ansatz 
i.e. $\sigma^2_{XN}=\beta \sigma_{X\bar{X}}\sigma_{NN}$ which, using the fact
that cross sections scale as the square of the potential can give a rough 
idea about the magnitude of $V_{XN}$ (as we do in the second part). 

In this brief note, we present two new results: (A) the first uses an 
earlier 
investigation of primordial nucleosynthesis of (low Z) anomalous nuclei 
containing $X$ \cite{dicus} for the case of sufficiently strong $X-N$ 
interaction to place limits on the $X$ abundance for the case when the
annihilation cross section is assumed as in Ref.\cite{moh} and 
$V_{XN}$ is left arbitrary and 
(B) the second uses the above mentioned ``factorization'' ansatz for the 
annihilation cross section to determine the likelihood of relic $X$s
binding in high Z nuclei for weaker $X-N$ interaction (Binding is more
likely for high Z nuclei since their larger radii imply lower kinetic 
energies.).

\noindent{\it A. Binding in low Z nuclei}

In Reference\cite{dicus}, Dicus and Teplitz calculated the abundance of 
anomalous
$Z~>~1$ nuclei that would be produced by primordial nucleosynthesis if there
existed a new neutral, stable, massive baryon $X^0$ ( the name used in 
\cite{dicus} was $H$). This was motivated by
 earlier work of Dover, Gaisser and Steigman\cite{dover} on
cosmological consequences of such particles. These papers assumed
that the lightest color singlet bound state involving the massive colored
particle and/or quarks and gluons interacts with nucleons in a similar
way to that of the $\Lambda$ hyperon

 The result of 
Reference \cite{dicus} was an estimate of the abundances of anomalous nuclei
relative to the total present abundance of all isotopes of $He$, $Li$,
$Be$, and $B$ for the case of low Z binding. It showed a 
significant enhancement in the abundance of $^9Be^*$. 
 Although $^8Be$ is not stable, the $\Lambda$
hyperfragment $^9Be^*_{\Lambda}$ ($^8Be$ to which a $\Lambda$ has been added)
is stable. Reference \cite{dicus} showed that replacing $\Lambda$ by another
 strongly interacting particle, such as $X^0$, present with a cosmic 
abundance $C_X\equiv \frac{n_X}{n_B}=\frac{\eta_X}{\eta_B}$ at cosmic 
nucleosynthesis, would
lead to a value for the ratio of the abundance $^9Be^*$
(anomalous $^9Be$) to the sum of all non-anomalous $Be$ isotopes observed 
today greater than the ratio of the abundances of $^5He^*$ to $^4He$ 
by a factor of about $10^{4}$.

We make the natural
assumption that any primordially synthesized anomalous heavy isotope
should be present in random samples on earth in the same abundance
as at the time of nucleosynthesis. However to be 
conservative, we will assume that 90\% of any primordial $^9Be^*$ will
have been destroyed in stars. This is a conservative estimate because
less than 90\% of the Universe's Hydrogen has been cycled through stars; 
indeed interstellar deuterium (deuterium is also astrated) is of the 
order of one third the value of primordial calculations.

We consider (i) the question of what $X-Nucleus$ (also 
denoted by $X-A$) cross section we expect on the basis of our knowledge 
of the $\Lambda$ hyperfragment and (ii) whether for such cross 
sections, $X^0$'s will be captured during cosmic nucleosynthesis.
We then apply the constraint of Ref.\cite{dicus} to limit either the $X-A$
cross section or the $X$ abundance in the light of the data of Ref.s
\cite{middle,middle2,hemmick}.

Following the line of reasoning given for $\Lambda$ hyperfragments by Z. 
Povh\cite{povh}, we use for the binding energy of $X$-particles in nuclei:
\begin{eqnarray}
B_{X}= V_{X} - \frac{\pi^2}{2\mu R^2}
\end{eqnarray}
where $V_X$ is the depth ($A$-independent) of the potential well for the 
$X-A$ system, $R=R_0(A-1)^{1/3}$, and $\mu$ is the reduced 
mass for the $X-A$ system. We note here that $X-N$ binding has been 
studied in a very helpful paper by R. Plaga\cite{plaga} for the case of $M_X 
< 1$ GeV. In the $\Lambda$ case, $R_0$ is 
about $1.5~Fermi$ and $V_{\Lambda}\simeq 27$ MeV. $B(^5He_{\Lambda})$
is about 3.1 MeV. Note that because of the absence of the one pion exchange
force from the $\Lambda-N$ potential, there is no $\Lambda$ analog 
to the deuteron ($^2H_{\Lambda}$) although $^3H_{\Lambda}$ is  bound.
Based on Eq. (2), we would expect $^5He^*$ which is crucial to the
formation of $Be^*$ and other light anomalous nuclei, to exist at the
time of nucleosynthesis for some other strongly interacting particle $X$
provided the $X-N$ interaction is strong enough to produce an effective 
square well of depth of 10.5 MeV. Here we have required that
the binding energy $B_X$ be greater than $2.2$ MeV because
primordial nucleosynthesis proceeds rapidly after the temperature becomes low
enough that the high energy tail of the $\gamma$ distribution can no longer
dissociate deuterium. The low energy (triplet) N-N cross section is about 4
barns whereas the low energy $\Lambda-N$ cross section is about $0.5$ barns.
Since the cross section goes as the square of the scattering amplitude which
is proportional to the effective potential in some approximation, this
implies if $\sigma(XN)$ is bigger than 0.1 barns, we expect to have a bound
state. Although we cannot be certain about the modifications to Eq.(2) 
due to large $M_X$, it works well for hyperfragments and our intuition is 
that as the kinetic energy decreases relative to the potential due to larger 
$M_X$ the approximation should improve.

We turn next to the question of whether the $X$ particle will be captured by
$^4He$ nuclei to form $^5He^*$ nuclei. The probability of capture is roughly
\begin{eqnarray}
P=n_{\gamma 0} \eta_B 0.1(T/T_0)^3 vt 10^{-24}\sigma_{b}
\end{eqnarray}
where $n_{\gamma 0}$ is today's CBR density; $\eta_B$ is the baryon to
photon ratio; the factor .1 takes into account account the Helium fraction 
relative
to the nucleons; T and t are the temperature and time at the time of
nucleosynthesis; $T_0$ is the CBR temperature today; and the relative 
velocity $v\simeq v_{He}\simeq 
(6T m^{-1}_{He})^{1/2}\simeq .1 c$ ($c$ is the velocity of light). Finally 
$\sigma_b$ is the cross section in barns for $X$ to be captured by $^4He$. 
We compute $\sigma_b$ following Blatt and Weiskopf\cite{bw} on the electric 
dipole moment contribution to $n+p\rightarrow d+\gamma$ as discussed in 
Plaga\cite{plaga}. For
$X$-capture by a nucleus of atomic weight A and atomic number Z. we have
\begin{eqnarray}
\sigma_{cap}(A, Z, k)\simeq 
\frac{8\pi\alpha 
Z^2}{3}\gamma^{-2}\left(\frac{k\gamma}{k^2+\gamma^2}\right)^3 
\left(\frac{\kappa}{k}\right)^2
\end{eqnarray}
where $\gamma=\sqrt{(2\mu B)}$, $k^2/2M_X\simeq .1$ MeV  
and $\kappa\simeq B$. Assuming $B\simeq 2.2 $ MeV, we estimate that
$\sigma_b\simeq 3\times 10^{-8}$ barns giving $P\sim 10^{-3}$ for large 
$M_X$ (even for $M_X\simeq M_N$, $P\simeq 0.1$). Note that this result 
corrects the assumption of Ref.\cite{dover} copied by Ref.\cite{dicus} that
$P\geq 1$. It shows that cosmic nucleosynthesis will not ``hide" stable
SIMPs in light nuclei so that signals from $X\bar{X}$ annihilation 
discussed in Ref. [5] and elswhere must be expected if there is no large 
$X-\bar{X}$ asymmetry.

Let us now consider the limits on the abundance of the $X$ if
the $X-N$ interaction is strong enough for formation of
$^5He^*$ and $^9Be^*$. To derive the abundance of $^9Be^*$, we need to know the
abundance of $^5He^*$ relative to normal $^4He$ at the time of 
nucleosynthesis since the reaction responsible for the synthesis of $^9Be^*$
is $^4He(^5He^*,\gamma)^9Be^*$. One can then write
\begin{eqnarray}
\frac{n_{^9Be^*}}{n_{Be}}|_{now}= \left(\frac{n_{^9Be^*}}{n_{^5He^*}}\right)
\left(\frac{n_{^5He^*}}{n_B}\right)\left(\frac{n_B}{n_{Be}}\right)
\end{eqnarray}
From Ref.\cite{dicus}, we get $^9Be^*$ to $^5He^*$ ratio to be $10^{-6}$;
Eq. (3) tells us that the second factor in the above equation is 
$10^{-3}C_X$,
and finally the present $Be$ abundance is taken fron Reeves\cite{reeves}
who gives $\frac{n_{Be}}{n_B}\simeq 2\times 10^{-11}$. Combining gives
the ratio $^9Be^*/Be\simeq 50C_X$. Other anomalous
isotopes would have abundances below $C_X$ according to Ref.\cite{dicus}.
In order to reach these conclusions, Ref.\cite{dicus} uses known rates for
analog production and depletion reactions.

The data on $Be$ is summarized by Hemmick et al.\cite{hemmick} as follows: 
Klein et al \cite{middle2} have ruled out isotopes of $Be$ up to 93 AMU at 
concentrations above $10^{-12}$ per nucleon using accelerator mass 
spectrometry whereas Hemmick et al limit the abundance to $10^{-11}$ or less
per nucleon for 100 to 1000 AMU and to $10^{-9}$ or less up to 10,000 AMU
\footnote{Profs. Elmore and Hemmick have assured us that the results of 
Ref.\cite{hemmick} obtain down to 93 AMU}.
 It should also be noted that Vanderschiff et al.\cite{van}
have placed limits on anomalous $^5He^*$ for $42\leq M_X\leq 82$.

Taking into account the above result that $Be^*/Be\simeq
50 C_X$ and assuming 90\% destruction of $^9Be^*$ in 
stars means that the limits on $C_X$ are: 
$3\times 10^{-13}$ to $M_X\leq 93$ GeV, $3\times 10^{-12}$ for $100 \leq 
M_X/GeV\leq 1000$ and $3\times 10^{-9}$ for $1000\leq M_X/GeV \leq 5000$ 
 and $3\times 10^{-8}$ between $5 \leq M_X/TeV \leq 10$. These limits
will apply to models in which a strongly interacting particle $X$ has
an attractive interaction with nucleons with sufficient strength that
the anomalous $Be^*$ forms with binding energy greater than 2.2 MeV.
Roughly such models should include cases in which $X-N$ scattering is 
attractive and $\sigma_{XN}\geq .1$ barns\footnote{Note that 
 if we used the ``factorization'' formula for cross sections and
$V_{XN}\sim 10$ MeV, we would obtain a much higher annihilation 
rate in the early universe and consequently a lower relic density $C_X$; 
nevertheless, it would still be inconflict with the bounds of 
Ref.\cite{hemmick}, thus maintaining our conclusion. } .

\noindent{\it B. Binding in high Z nuclei}

We now consider the case in which $V_{XN}$ is too small for binding
with $He$ or $Be$. Even though we do not know the nature of the $X-N$ force 
from
fundamental principles, we can put a lower bound on $V_{XN}$ using the 
afore mentioned ``factorization'' hypothesis i.e. $\sigma^2_{XN}=\beta 
\sigma_{X\bar{X}}\sigma_{NN}$  which gives
\begin{eqnarray}
V_{XN}\sim V_{NN}(\sigma_{XN}/\sigma_{NN})^{1/2}\nonumber\\
> V_{NN}[(\sigma^{min}_{X\bar{X}}\sigma_{NN})^{1/2}/
\sigma_{NN}]^{1/2}\beta^{1/4}
\end{eqnarray}
For $\sigma^{min}_{X\bar{X}}$, we use the fact that $X\bar{X}$ annihilation
cross-section must be sufficiently large to ensure that today $M_Xn_X\leq 
\rho_c$. Using standard methods\cite{kolb}, we find $\sigma_{X-\bar{X}}
\geq 3\times 10^{-13}$ barns (independent of $M_X$) which implies 
$V_{XN}\simeq 20$ KeV if $V_{NN}\sim 50$ MeV. Eq. (2) then gives a 
condition for the existence of the bound state 
(in the approximation $B_X\ll V_X$):
\begin{eqnarray}
0.425 A^{2/3}\beta^{-1/4}\frac{V_{XN}}{V_{NN}}\geq A^{-1}+M^{-1}_X
\end{eqnarray}
where by $M_X$ we mean $M_X$ in GeV's. For $A\ll (M_X/GeV)$, this yields 
$A^{5/3}\geq 2.3\frac{V_{NN}}{V_{XN}}\beta^{1/4}$. This yields 
$A\simeq 183\beta^{3/20}$ for $R_0=1.3~Fermi$ and $\simeq 154\beta^{3/20}$
for $R_0=1.5~Fermi$. For the case where $A\gg (M_X/GeV)$, we have instead the
condition for bound state formation $A^{2/3}\geq \frac{\pi^2}{2 M_X R^2_0}$
which yields, $A\geq 427$ for $M_X=10$ GeV.
Bound states may exist for smaller $A$ and/or $M_X$ if 
$V_{XN}$ 
exceeds its minimum. It is important to note that the lower bound on $A$ 
is very insensitive to the value $\beta$ which characterizes any 
deviation from the formula $\sigma^2_{XN}=\sigma_{NN}\sigma_{X\bar{X}}$. 
If we set $A=238$ in Eq. (8), we should be able to detect bound states 
of $X$'s with minimal interaction down to $M_X=83$ GeV.

We can again use Eq. (5) to compute the rate for 
$X$-capture. Assuming that $X$-particles have the galactic virial veloclty
($v\sim 10^{-3}c$) and halo dark matter density ($\rho_H= 5\times 10^{-25}~
gm~cm^{-3}$), we find for Unranium $\sigma\sim 10^{-8}$ barns. The 
probability of $U$ capturing an $X$ in the lifetime of the solar
system is then $P=\frac{\rho_H}{M_X}\sigma v_{vir}\tau_{solar}\simeq
1.6\times 10^{-11}/(M_X(TeV))$.

One can extend the above line of reasoning to discuss what happens when
the $X$ particles have nuclear cross sections necessary for them to play 
the role of UHECR's. Consider for example the case discussed in Ref. 3 where
they need to have a $\sigma_{XN}$ of order of a milli-barn. Using our 
simple scaling laws, we conclude that this implies  $V_{XN}\simeq 
10^{-2} V_{NN}$. Inserting in Eq. (2) gives for large $M_X$ that
bound states should exist for $A\geq 30$ (for $A\ll (M_X/GeV)$) assuming of 
course that the potential is attractive. Setting $A=238$ gives bound 
states for $M_X > 6.3$ GeV. In this case, the capture cross section ranges 
from $10^{-4}$ b for $M_X=6.3$ GeV to $10^{-7}$ b for large $M_X$ so that 
abundances are larger than minimal interaction case (assuming similar 
galactic halo dark matter).

To date searches for anomalous isotopes only seem to have reached $Z=9$ 
(Fluorine). Based on the above considerations, we urge experimentalists to 
search for anomalous isotopes with highest Z so that different interaction 
strengths for SIMPs may be explored. In particular, anomalous Uranium 
may provide a higher discovery potential if the halo
dark matter is in fact dominated by ``minimally'' strongly interacting 
massive particles. However searches in much lighter elements $A\geq 30$
should stable SIMPs capable of explaining UHECRs.

 \bigskip
\noindent{\Large \bf{Acknowledgements}}
\bigskip

The work of R. N. M.
is supported by the National Science Foundation grant under no. PHY-9802551
and the work of V. L. T. is supported by the DOE  under grant no.
DE-FG03-95ER40908. We would like to thank R. Boyd, S. Brodsky, D. A. 
Dicus, D. Elmore, G. Farrar, E. Fishbach, T. Hemmick, Xiangdong Ji, E. W. 
Kolb, R. Middleton, G. Steigman and S. Wallace for discussions.

\end{document}